\documentclass[preprint]{aastex}

\slugcomment{accepted to AJ}
\shorttitle{Uncertainty in the Orbits of Extrasolar Planets}
\shortauthors{Ford}
\begin{document}
\title{Quantifying the Uncertainty in the Orbits of Extrasolar Planets}
\author{Eric B.\ Ford}

\affil{Department of Astrophysical Sciences, 
	Princeton University, 
	Peyton Hall, 
	Princeton, NJ 08544-1001, USA}

\altaffiltext{1}{present address: Astronomy Department, 
	601 Campbell Hall, 
	University of California at Berkeley, 
	Berkeley, CA 94720-3411, USA}
\email{eford@astron.berkeley.edu}

\begin{abstract}
Precise radial velocity measurements have led to the
discovery of $\sim100$ extrasolar planetary systems.  We investigate
the uncertainty in the orbital solutions that have been fit to these
observations.  Understanding these uncertainties will become more and
more important as the discovery space for extrasolar planets shifts to
longer and longer periods.  While detections of short period planets
can be rapidly refined, planets with long orbital periods will require
observations spanning decades to constrain the orbital parameters
precisely.  Already in some cases, multiple distinct orbital solutions
provide similarly good fits, particularly in multiple planet systems.
We present a method for quantifying the uncertainties in orbital fits
and addressing specific questions directly from the observational data
rather than relying on best fit orbital solutions.  This Markov chain
Monte Carlo (MCMC) technique has the advantage that it is well suited
for the high dimensional parameter spaces necessary for the multiple
planet systems.  We apply the MCMC technique to several extrasolar
planetary systems, assessing the uncertainties in orbital elements for
several systems.  Our MCMC simulations demonstrate that for some
systems there are strong correlations between orbital parameters
and/or significant non-Gaussianities in parameter distributions, even
though the measurement errors are nearly Gaussian.  Once these effects are
considered the actual uncertainties in orbital elements can be
significantly larger or smaller than the published uncertainties.  We
also present simple applications of our methods such as predicting the
times of possible transits for GJ 876.
\end{abstract}


\section{Introduction}

Recent detections of planets around other stars have spurred a
wide range of research on planet formation and planetary system
evolution.  The first planet discovered around a solar type star, 51
Pegasi b, was in a surprisingly short period orbit \cite{Mayor95}.
Other early planets such as 70 Virginis b revealed surprisingly large
orbital eccentricities \cite{MaBu96}.  Multiple planet systems have
revealed intricate dynamical interactions such as the 
resonances in Upsilon Andromedae \cite{BuMaFi99} and GJ876 \cite{mbv2001b}.

The future of radial velocity planet searches promises to be exciting.
Ongoing large surveys including a broad array of nearby main-sequence
stars will continue to increase the number of known extrasolar
planets.  Refinements in the radial velocity technique should continue to
improve measurement precision, permitting the detection of planets
with smaller masses.  The increasing
time span of precision observations will permit the discovery of
planets with larger orbital periods.  Several stars presently known to
harbor one planet are expected to reveal additional planets in longer
period orbits \cite{Fischer01}.

These advances will also bring new challenges.  The 51 Pegasi-like
planets could be rapidly confirmed by independent observers
\cite{Mayor95,Marcy97} due to their short orbital periods and large
velocity amplitudes.  However, for a given signal-to-noise ratio, the
time required to obtain observations to confirm or refute a possible
planetary candidate will typically scale with the orbital period of
the planet.  Already, one known planet (55 Cancri d) has an orbital
period of over 14 years \cite{mbflvhp2002}.

It is also more difficult to obtain precise orbital elements for
planets with large orbital periods.  While early planet candidates
were routinely observed for multiple periods before publication,
recently planet candidates have been published when observations span
only a single orbital period.  Thus, it will become increasingly easy
to over-interpret early orbital determinations.  A further challenge
is that the growing precision and time span of radial-velocity
measurements are expected to significantly increase the number of stars
known to harbor multiple planets.  Fitting multiple-planet systems
requires many more free parameters, so that observational data may not
tightly constrain all the orbital parameters or even distinguish
between multiple possible orbital solutions, especially in the
presence of significant noise and sparsely sampled data.

These trends imply that it will become increasingly important to
understand the uncertainties in orbital elements and other parameters
derived from such observations, and thus we must use the best possible
statistical tools to analyze radial velocity data.  

In this paper we introduce a Bayesian analysis for constraining the
orbital parameters of extrasolar planets with radial velocity
observations.
The Bayesian framework considers a joint probability distribution
function for both the observed data ($\vec{d}$) and model parameters
which can not be directly observed ($\vec{x}$).  This joint
probability, $p(\vec{d}, \vec{x})$, can be expressed as the product of
the probability of the observables given the model parameters,
$p(\vec{d} | \vec{x})$, and a prior probability distribution function,
$p(\vec{x})$, which is based on previous knowledge of the model
parameters.  Bayes's theorem allows one to compute a posterior
probability density function, $p(\vec{x} | \vec{d})$, which incorporates
the knowledge gained by the observations $\vec{d}$.  That is
\begin{equation}
p(\vec{x}| \vec{d}) 
= \frac{ p( \vec{d}, \vec{x}) }{ \int p( \vec{d}, \vec{x}) p( \vec{x}) \,d\vec{x} }
= \frac{ p( \vec{x}) p(\vec{d} | \vec{x}) }{ \int p( \vec{d}, \vec{x}) p( \vec{x}) \,d\vec{x} }.
\end{equation}
Unfortunately, the lower integral can be extremely difficult to
compute, particularly when $\vec{x}$ has a large number of dimensions.
Even after this integral is performed, most questions require
additional integration of $p(\vec{x}|\vec{d})$ over many of the model
parameters.  This paper describes the application of Markov chain
Monte Carlo (MCMC) simulation using the Metropolis-Hastings algorithm
and the Gibbs sampler to perform the necessary integrations.  This
technique allows us to accurately characterize the posterior
probability distribution function for orbital parameters based on
radial velocity observations.

In this paper we first describe our model for calculating radial
velocities from planetary orbital parameters.  In \S3 we summarize
methods for identifying the maximum likelihood orbital solution for a
set of observed radial velocities.  Then, we discuss methods of
characterizing the uncertainties of model parameters in \S 4.  We
focus our attention on the application of Markov chain Monte Carlo
(MCMC) simulation to estimate the uncertainties for orbital parameters
in a Bayesian framework.  In \S 5 we apply this technique to analyze
several published extrasolar planet data sets.  Finally, we summarize
the potential of this technique as additional long period planets and
multiple planet systems are discovered.

\section{Radial Velocity Model}

In radial velocity surveys, the velocity of the central star is precisely
monitored for periodic variations which could be caused by orbiting
companions.  Each individual observation can be reduced to a
measurement of the star's radial velocity and an estimate of the
observational uncertainty based on photon statistics.  Because the
observations are averaged over hundreds of sections of the spectrum,
the observational uncertainties of most current echelle based radial
velocity surveys are nearly Gaussian \cite{BuMa98}.  Stellar
activity can also contribute to the observed radial velocities.
However, most of the stars
investigated in this paper are chromospherically quiet and so that the
contribution of stellar ``jitter'' to the observed radial velocities
is believed to be negligible.

\subsection{Single Planet Systems}

Such radial velocity observations are able to detect the acceleration of a
star due to the gravitational perturbations due to an orbiting planet.
We model the motion of the planet as a Keplerian orbit using the
parameters: orbital period ($P$), velocity semi-amplitude ($K$),
eccentricity ($e$), argument of periastron ($\omega$), and mean
anomaly at the specified epoch ($M_o$).  The perturbation to the
radial velocity ($\Delta v_*$) of a star due to a planet on a Keplerian
orbit is given by
\begin{equation}
\label{rveqn}
\Delta v_*(t) = K \left[ \cos\left(\omega+T(t)\right) + e \cos(\omega) \right] 
\end{equation}
A planet's true anomaly ($T(t)$) is a function of time ($t$) and is
related to the planet's eccentric anomaly ($E(t)$) via the relation
\begin{equation}
\label{true anomaly}
\arctan \left( \frac{T(t)}{2} \right) = \sqrt{\frac{1+e}{1-e}} \arctan \left( \frac{E(t)}{2} \right).
\end{equation}
The eccentric anomaly is related to the mean anomaly ($M(t)$) via Kepler's equation
\begin{equation}
\label{keplereqn}
E(t) - e \sin\left(E(t)\right) = M(t) - M_o = \frac{2\pi}{P} \left( t - t_o \right)
\end{equation}
where $M_o$ is a constant, the orbital phase at $t=0$ which is related to the time of pericenter ($t_o$).  

\subsection{Multiple Planets}

For a star being perturbed by multiple planets, there is no analytic
expression for the exact radial velocity perturbation.  However, in
many cases the radial velocity perturbation can be well modeled as the
sum of multiple independent Keplerian orbits.  The deviations from the
simple sum of Keplerians model can be divided into two types:
short-period interactions and secular interactions.  The magnitude of
the short-period interactions (deviations from independent Keplerian orbits on an
orbital timescale) is often small compared to the magnitude of the
Keplerian perturbation and the observational uncertainties.  The
secular interactions are typically modeled as changes in the Keplerian
orbital parameters.  While the secular interactions can have large
effects on the observed radial velocities, the timescales are
typically much longer than the time span of observations.  Thus, we
model the observed radial velocity of a star as

\begin{equation}
\label{multiplaneteqn}
v_{*,\rm model}(t,j) = \sum_i \Delta v_{*,i}(t) + C_j
\end{equation}
where $\Delta v_{*,i}(t)$ is given by Eqn.\ \ref{rveqn} using the orbital
parameters of the $i$th planet and there is an unknown constant
velocity offset ($C_j$) which provides no information about planetary
companions.  In practice, high precision radial velocity surveys are
typically calibrated such that different observatories have different
zero-point offsets, and hence it is important to allow 
observations from different observatories to have independent constant
velocity offsets.

Since the uncertainties from individual observational are expected to
closely follow a normal distribution, To evaluate the goodness of fit
for a given model, we calculate the usual $\chi^2(\vec{x})$ statistic,
\begin{equation}
\label{chisqeqn}
\chi^2 = \sum_k \left( v_{*,\rm model}\left(t_k,j_k\right) - v_{*,\rm obs}\left(t_k,j_k\right) \right)^2 / \sigma_{k}^2
\end{equation}
where $t_k$ is the time of the $k$th observation, $j_k$ is the
observatory used for the $k$th observation, and $\sigma_k$ is
the observational uncertainty of the $k$th radial velocity
observation.

\section{Maximum Likelihood Estimators}

Given a set of model parameters ($\vec{x}$), which are typically the
masses and orbital parameters of the planets, we wish to compare the
model predictions to the observations using the model and
$\chi^2(\vec{x})$ statistic described in \S2.  The maximum likelihood
estimate (MLE) of the model parameters $\vec{x}_{MLE}$ is obtained by
finding the minimum $\chi^2(\vec{x}_{MLE}) = \min_{\vec{x}} \chi^2$.
These are the orbital parameters typically reported as the
``best-fit'' model.  Finding the set of parameters which minimizes
$\chi^2(\vec{x})$ can be challenging, particularly for multiple planet
systems.  Often a combination of several methods is used to identify
the ``best-fit'' orbital parameters.

\subsection{Periodograms}

Promising orbital periods can be
recognized as sharp dips in a plot of the period versus the minimum
$\chi^2$ for that period with the phase and amplitude allowed to
vary (a periodogram).  There are well developed methods for determining the
significance of periodicities identified in this way and estimating the false
alarm probability \cite{HorneBaliunas86}.  While a single sinusoid
cannot reproduce the signal of an eccentric Keplerian orbit,
a periodogram allows the rapid identification of any periodicities in
observational data, without requiring a simultaneous fit for
amplitude, eccentricity, argument of periastron, or other parameters.
While the use of periodograms for an initial exploration of parameter
space may not be optimal for identifying all possible Keplerian
variations in the radial velocities (e.g. large eccentricities spread
power across multiple frequencies), they do provide an efficient means
for identifying potential orbital periods.  Accurate estimates of the
orbital period are valuable input parameters for subsequent
algorithms, particularly the local minimization procedures.

\subsection{Local Minimization}

Once an initial estimate of the orbital parameters is available
($\vec{x}_o)$, an iterative minimization algorithm such as
Levenberg-Marquardt (LM) \cite{NR} can be used to refine the model
parameters by minimizing $\chi^2$.  The ML algorithm will identify
only a single local minimum for a given initial guess of model
parameters, $\vec{x}_o$.  Unfortunately, the $\chi^2(\vec{x})$ surface
can have many local minima.  In particular, for multiple planet
systems, the many degrees of freedom and the rugged $\chi^2(\vec{x})$
surface can render LM minimization particularly sensitive to the
initial guess model parameters and vulnerable to finding a local
minimum of $\chi^2(\vec{x})$ far away from and much less probable than
the global minimum.  In practice, local minimization algorithm such as
LM are most useful for fine tuning the parameters of a local minima
identified by global search algorithms.

\subsection{Simulated Annealing}

Simulated annealing generalizes iterative algorithms such as LM to
reduce the risk of becoming trapped in a local minimum \cite {NR}.
Random perturbations are applied at each iteration.  The perturbations
are initially large (high temperature) and are gradually reduced
(lower temperature).  Provided the temperature is reduced
sufficiently slowly, simulated annealing can convert a local
minimization algorithm into a global minimization algorithm.
Unfortunately, the choice of the cooling curve is
important.  Additionally, the number of iterations required may be
prohibitively large.  Nevertheless, simulated annealing can be a
valuable tool when performing global non-linear minimization.

\subsection{Genetic Algorithms}

A genetic algorithm (GA) is an optimization algorithm loosely based on
biological evolution \cite{Charbonneau95}.  GAs are much less likely
to become trapped in local minima than local minimization algorithms
such as LM.  The disadvantage is that they require orders of magnitude
more evaluations of the goodness of fit statistic. When fitting
planetary orbits to a single set of radial velocity data, this can be
merely an inconvenient delay for analytic models, but GAs can be
prohibitively time consuming if it is necessary to search a large
number of data sets.  Once a GA has identified a minimum, a LM-type
minimization algorithm can provide an efficient means of fine tuning
the solution.  GAs have been applied to the $\upsilon$ And and GJ876
systems \cite{StepMalhBla2000,LaughCamb2001}.  In these cases the GAs
have verified the minima found by previous authors.  Additional minima
have been identified in the case of $\upsilon$ And
\cite{StepMalhBla2000}.

\section{Algorithms for Estimating Uncertainties}

The previous discussion has focused on identifying the maximum
likelihood or best-fit parameter values.  It is also important to
characterize the uncertainty in the estimation of the parameter
values.  Here we discuss three methods of estimating these
uncertainties: constant $\Delta\chi^2$ boundaries, resampling, and
Markov chain Monte Carlo (MCMC) methods.

\subsection{Constant $\Delta\chi^2$ Boundaries}

One method of estimating the uncertainty in orbital parameters fit to
the observed data is to evaluate $\chi^2(\vec{x})$ at points on a grid
in the parameter space near $\vec{x}_{MLE}$.  Then confidence
intervals can be calculated by finding the boundary (B) along which
$\chi^2$ is constant.  The probability, $P$, that the parameters lie inside
the boundary can be calculated by 
\begin{equation}
\frac{\int_B\, dx e^{-\Delta\chi^2(\vec{x})} }
{\int\, dx e^{-\Delta\chi^2(\vec{x})}},
\end{equation}
where $\Delta\chi^2(\vec{x}) = \chi^2(\vec{x})-\chi^2(\vec{x}_{MLE})$,
and the bottom integral extends over the entire parameter space.  This
approach is equivalent to assuming a uniform prior in $\vec{x}$ in a
Bayesian framework.  Unfortunately, evaluating these integrals often
requires a prohibitive number of evaluations of $\chi^2(\vec{x})$,
particularly when the number of parameters is large.  For example, to
fit the radial velocities caused by one, two, or three planets
requires at least six, eleven, or sixteen model parameters,
respectively.  If only ten grid points are used in each dimension,
then this technique would require $10^6$, $10^{11}$, or $10^{16}$
evaluations of $\chi^2(\vec{x})$.  In practice, the integrands are can
be poorly behaved, requiring many more evaluations in each dimension.
Thus, this technique is usually not practical for estimating the
uncertainties in orbital parameters, especially for multiple planet
systems.

Brown (2004) has applied a similar method to the case of HD 72659.
Brown evaluated $\chi^2$ choosing random parameter values within a
region of parameter space using Monte Carlo rather than choosing
parameters along a grid.  Still, the large range of parameter space
made it impractical to explore the entire range of parameter space
with a constant density.  Thus, Brown initially sampled a wide range
of parameter space and then manually identified several boxes of
parameters space to sample at higher density.  In total Brown
evaluated $\chi^2$ at hundreds of millions of points to 
sample the allowed parameter space for HD 72659.

\subsection{Refitting to Synthetic Data Sets}

One method of estimating the uncertainty in orbital parameters fit to
the observed data is to apply the fitting technique repeatedly to many
sets of simulated data.  Since the observation errors are believed to
be very nearly Gaussian and each radial velocity measurement has a
corresponding uncertainty estimate, it is straightforward to construct
simulated data sets by adding Gaussian random values to the actual
data points.  Each set of simulated data is meant to represent a
possible set of measurement values.  If the same fitting procedure is
applied to the actual data and each simulated data set, then one can
obtain the distribution of best-fit parameter values.  

One disadvantage of the refitting technique is that one must identify
the best-fit orbital parameters for each synthetic data set.  In
principle, one should apply a global minimization algorithm to each
data set, but in practice the computational requirements often dictate
that only a local minimization routine will be run for the synthetic
data sets.  Even when using a local minimization algorithm, the
computational requirements can be a burden.  For example, for a one
planet system with six free parameters, $\chi^2$ must be calculated
thirteen times for each iteration of LM for each synthetic data set.
For obtaining confidence intervals roughly equivalent to 3-$\sigma$, a
sample of at least $\sim10^5$ synthetic data sets is necessary.
Assuming that the local minimization routine requires an average of
eight iterations to converge, this amounts to $\sim 10^7$ evaluations
of $\chi^2$.  Still, the use of only a local minimization algorithm on
the synthetic data sets would render this type of analysis vulnerable
to underestimating the range of allowed parameters.  If a global
minimization algorithm were used, even over a small fraction of the
possible parameter space, the number of $\chi^2$ evaluations required
would increase by orders of magnitude.

Unfortunately, even then the distribution estimated via resampling may
not reflect the full range of possible parameter values, particularly
in cases where the $\chi^2$ surface is significantly asymmetric around
the minimum.  To illustrate this possibility, we use the actual
observations for the extrasolar planet around HD 72659.  In Figure 1,
we show the best-fit value of $\chi^2$ when all parameters except
orbital period are allowed to vary as a function of the orbital
period (solid line).  We also show with dotted lines the same curve,
but using synthetic data sets generated from the actual observations
of HD 72659.  For all the data sets, orbital periods shorter than $\sim 1700$d
are strongly ruled out and $\chi^2$ increases very slowly for orbital
periods greater than the best-fit period.  However, the variation in
the location of the best-fit period across the synthetic data sets
does not reflect the fact that the very slow increase in $\chi^2$ for
longer period orbits allows a very large range of orbital periods.

\subsection{Markov chain Monte Carlo, the Metropolis-Hastings Algorithm, \& the Gibbs Sampler}

Bayesian inference using Markov chain Monte Carlo (MCMC) simulations
provides an alternative method for estimating the uncertainty of
fitted parameters.
The MCMC method has been applied to several other astronomical data
sets and problems, including spectra analysis (Kashyap \& Drake 1998,
van Dyk et al.~2001), star formation history (Fernandes et al.~2001,
Panter, Heavens, \& Jimenez 2003), object detection (Hobson \&
McLachlan 2003), the Cepheid distance scale (Barnes et al.~2003),
cluster weak lensing (Rodriguez 2003), the SZ effect (Marshall,
Hobson, \& Slosar 2003), type Ia supernovae (Wang, Yun, \& Mukherjee
2004), and especially the cosmic microwave background (e.g., Knox,
Nelson, \& Skordis 2001, Verde et al.~2003).

The goal of the MCMC method is to generate a chain (i.e. sequence) of
states (i.e. sets of parameter values, $\vec{x}_i$) which are sampled
from a desired probability distribution ($f(\vec{x})$).  Such a chain
can be calculated by specifying an initial set of parameter values,
$\vec{x}_0$, and a transition probability, $p(\vec{x}_{n+1} |
\vec{x}_n)$.  The Monte Carlo aspect of MCMC simulation refers to
randomness in the generation of each subsequent state.  The Markov
property specifies that the probability distribution for determination
of $\vec{x}_{n+1}$ can depend on $\vec{x}_n$, but not previous states.
If the Markov chain is reversible, that is, if
\begin{equation}
f(\vec{x}) p(\vec{x}|\vec{x}') = f(\vec{x}') p(\vec{x}'|\vec{x}),
\end{equation}
aperiodic, and irreducible, then it can be proved that the Markov
chain will eventually converge to the stationary distribution
$f(\vec{x})$ (Gilks, Richardson, \& Spiegelhalter 1996).  The
requirement that the chain be irreducible guarantees that it is
possible for the chain to reach every state with non-zero probability
from any initial state.

MCMC offers a relatively efficient method of performing the
integrations necessary for a Bayesian analysis (eqn.\ 1), if we can
calculate a Markov chain whose equilibrium distribution is equal to
the joint posterior probability density function for the model
parameters given the observed data

For application to radial velocity measurements and orbit
determination, the observational errors are believed to be very nearly
Gaussian with accurately estimated variances.  Thus, if the data were
generated by the model specified by $\vec{x}$, then the probability of
drawing the observed values, $p(\vec{d}|\vec{x})$, is proportional to
$\sim \exp\left(-\chi^2(\vec{x})/2\right)$.  If we choose a uniform
prior in $\vec{x}$ ($p(\vec{x}) \sim 1$), then the posterior
distribution, $p(\vec{x}|\vec{d})$, is also proportional to $\sim
\exp\left(-\chi^2(\vec{x})/2\right)$.  This will allow us to construct
Markov chains with an equilibrium distribution equal to the posterior
distribution,
\begin{equation}
f(\vec{x}) = p(\vec{x}|\vec{d}) = p(\vec{x}) p(\vec{d}|\vec{x}) \sim \exp\left(-\chi^2(\vec{x})/2\right).
\end{equation}

The Metropolis-Hastings (MH) algorithm involves the generation of a
trial state ($\vec{x}'$) according to a candidate transition
probability function, $q(\vec{x}'|\vec{x}_n)$, and randomly accepting
the trial as the next state or rejecting the trial state in favor of
the current state.  The MH algorithm specifies an acceptance
probability, $\alpha(\vec{x}'|\vec{x})$, such that the transition
probability
\begin{equation}
p(\vec{x}'|\vec{x}) = q(\vec{x}'|\vec{x}) \alpha(\vec{x}'|\vec{x})
\end{equation}
is guaranteed to be reversible and irreducible, provided only that
$q(\vec{x}'|\vec{x})$ allows transitions to all $\vec{x}$ for which
$f(\vec{x})$ is non-zero.  The MH algorithm acceptance probability is
\begin{equation}
\alpha(\vec{x}'|\vec{x}) = \min \left\{ \frac{f(\vec{x}')
q(\vec{x}|\vec{x}')}{f(\vec{x}) q(\vec{x}'|\vec{x}) }, 1 \right\}
\label{eqnAlpha}
\end{equation}
if $f(\vec{x}) q(\vec{x}'|\vec{x}) > 0$ and
$\alpha(\vec{x}'|\vec{x}_n) = 1$ otherwise.  Note that the MH
algorithm does not require that the normalization of $f(\vec{x})$ be
known {\em a priori}.  While the MH algorithm guarantees that the
chain will converge to $f(\vec{x})$, it does not specify when the
chain will achieve convergence.

In principle, the basic MH algorithm can be implemented as follows.  
\begin{enumerate}
\item Initialize the chain with some $\vec{x}_0$ and $n=0$.  
\item \label{McmcLoop} Generate a trial state, $\vec{x}'$, according to $q(\vec{x}'|\vec{x}_n)$.
\item Calculate $\chi^2(\vec{x}')$ for trial state and $\chi^2(\vec{x}_n)$ for the current state
\item Determine the ratio $f(\vec{x}') / f(\vec{x}_n) \sim \exp \left( -\left[\chi^2(\vec{x}')-\chi^2(\vec{x}_n) \right]/2\right)$.
\item Draw a random number, $u$, from a uniform distribution between zero and one.
\item If $u\le \alpha(\vec{x}'|\vec{x}_n)$, as defined by Eqn.\ (\ref{eqnAlpha}), then set
$\vec{x}_{n+1} = \vec{x}'$.  If $u>\alpha(\vec{x}'|\vec{x}_n)$, then
set $\vec{x}_{n+1} = \vec{x}_{n}$.  
\item Set $n = n+1$.
\item Go to step \#\ref{McmcLoop}. 
\end{enumerate}

The choice of $q(\vec{x}'|\vec{x})$ and deciding when a chain has
converged are the primary practical complications.  The MH algorithm
can be optimized for a particular problem by the judicious choice of
$q(\vec{x}'|\vec{x})$.  Poor choices can lead to extremely inefficient
sampling and hence slow convergence.  The most efficient choice for
$q(\vec{x}'|\vec{x})$ would be $p(\vec{x'} | \vec{d} )$, the posterior
probability distribution itself.  However, this is not possible for
this application, since the whole purpose of the Markov chain is to
calculate the posterior distribution.  In such cases, a common choice
for $q(\vec{x}'|\vec{x})$ is a Gaussian distribution centered around
$\vec{x}$.  Still, there remain important choices about the
correlations and scale of the candidate transition probability
distribution.  We will address correlations later in this section, but
first discuss the choice of scale.  If the trial states are chosen
with too large a dispersion then a large fraction of the trial states
will be rejected, causing the chain to remain at each state for
several trials and to converge very slowly.  If the trial states are
chosen with too small a dispersion ($\sigma_{q}$), then the small step
size will cause the chain will behave like a random walk, i.e., the
number of steps required to for the chain to traverse a distance, $L$,
in parameter space would scale roughly as $\sim L^2/\sigma_{q}^2$.

Monitoring the fraction of trial states that are accepted is one way to 
verify that the scale chosen for $q(\vec{x}'|\vec{x})$ is not too inefficient.  
Optimal values for the acceptance rate have been determined for some simple cases.  
For example, when $\vec{x}$ has only one dimension and the posterior distribution
function is known to be Gaussian, then an acceptance rate of $\sim 0.44$ is optimal,
among the class of Gaussian candidate transition probability distribution functions 
centered on the current value of $\vec{x}$.  For parameter spaces with many 
dimensions a similar analysis yields an optimal acceptance rate of $\sim 0.25$
(Gelman et al.\ 2003).  

Thus, it is common to choose a Gaussian $q(\vec{x}'|\vec{x})$ centered
on $\vec{x}$ with some guess for the scale parameters.  If, after
running the Markov chain for some time, it becomes clear that the
acceptance rate is significantly different than the desired acceptance
rate, then the scale parameters are adjusted, the previous Markov
chain is discarded, and a new Markov chain is begun.  It may be
necessary to repeat this several times to determine an acceptable set
of scale parameters.  Note that Markov chains which serve as the basis
for altering the step size are not combined with the final Markov chain
for inference.

For multidimensional parameter spaces it is not always obvious how to
change $q(\vec{x}'|\vec{x})$ so as to obtain the desired acceptance
rate.  Even if all the parameters are uncorrelated, there are multiple
possible scale parameters.  For this reason, we chose to use a special
case of the Metropolis-Hastings algorithm, widely known as the
Metropolis-Hasting algorithm within the Gibbs sampler.  The Gibbs
sampler generates a trial $\vec{x}'$ by altering only a subset of the
parameters in $\vec{x}$ for each step.  We combine the Gibbs sampler
with a Gaussian candidate transition function, i.e., for the
parameters in ($\vec{x}$) to be updated, the candidate transition
probability function is
\begin{equation}
q(x'_\mu | x_\mu) = \frac{1}{\sqrt{2\pi \beta_\mu^2}} \exp \left[ -\frac{(x'_\mu-x_\mu)^2}{2\beta_\mu^2} \right]
\label{eqnCandTransProb}
\end{equation}
for valid $x'_\mu$  (i.e., if the model dictates that $x'_\mu$ be positive definite, then trial states with negative $x'_{\mu}$ are rejected).  
We use the index $\mu$ to distinguish
elements of the vector of parameters, and the index $n$ to indicate the
$n$-th step of the Markov chain.  Each $\beta_\mu$ is a parameter which controls the size
of the steps for the parameter indicated by $\mu$.  Thus, our acceptance probability reduces to
\begin{equation}
\alpha(\vec{x}'|\vec{x}) = \min \left\{\exp 
\left(\frac{\chi^2(\vec{x})-\chi^2(\vec{x}')}{2}\right),1\right\}
\end{equation}
for valid $\vec{x}'$, and $\alpha(\vec{x}'|\vec{x}_n) = 0$ for invalid
parameter values.

While there are several variations, we choose the parameters to be
altered in the next trial state according to randomly generated
permutations of the model parameters.  If the parameters are chosen
one at at time, then it is easy to monitor the acceptance rates for
steps involving each parameter separately.  If any of these acceptance
rates differ from the desired acceptance rate, then the guilty scale
parameters, $\beta_\mu$, can be identified and adjusted for
calculating the next Markov chain more efficiently.

The other major difficulty in applying MCMC is determining how long
the Markov chain should be before using it for inference.  
In practice, one performs
multiple tests which can positively identify chains which have not
converged.  The failure of such tests to demonstrate non-convergence
suggests, but does not prove, that the chain has converged \cite{Chen2002}.  
In our experience, we have found it advisable to construct multiple 
Markov chains for comparison.  If all the chains have converged, then 
the distributions of all quantities of interest should be similar to within 
statistical uncertainties \cite{GellmanRubin92}.

\subsection{Comparison}

Both refitting to simulated data sets and MCMC provide reasonable
methods of assessing the uncertainty in model parameters.  Refitting
to simulated data estimates the distribution of the MLE for the
simulated data, which is not necessarily the most desirable estimator.
Transcribed into the language of Bayesian statistics, the MLE is not
necessary the mean or median of the parameter's marginal distribution.
MCMC samples from the joint posterior distribution given a prior and
the observational data.  

While changing the prior would require
completely redoing an MLE analysis, importance sampling can be applied
to a realization of a Markov chain to compute confidence intervals for
several different priors simultaneously (Press et al.\ 1992; Gilks et
al.\ 1996).  This could become valuable when considering alternative
formation scenarios (e.g.  alternative distributions for initial
eccentricities and inclinations) or when additional constraints become
available (e.g. new observations or long term orbital stability
studies).  Note that for importance sampling to be accurate, the
posterior distributions under the different priors must not be too
dissimilar.

Further, calculating the best fit model for each set of simulated data
requires repeatedly solving a complex and time consuming non-linear
minimization problem.  Similarly, most techniques which are less
sensitive to local minima (e.g. GAs) are less efficient than MCMC for
multiple planet systems which have a large number of free parameters.
Calculating a longer Markov chain is a relatively simple task and
computationally more efficient.  Thus, we expect the MCMC technique to
generalize to multiple planet systems much better than refitting.

While both methods may be practical when fitting to a small number of
model parameters, MCMC is particularly useful for high dimensional
parameter spaces.  This advantage is particularly important for long
period planets and multiple planet systems where multiple free
parameters which can be traded off against each other to obtain
similarly good fits.  The radial velocity observations of such systems
can result in large valleys in the $\chi^2(\vec{x})$ surface which
permit a broad range of parameter values.  Within these valleys, there
is often a rugged $\chi^2(\vec{x})$ surface, meaning that there are
many nearby local minima for $\chi^2(\vec{x})$.  While the rugged
$\chi^2(\vec{x})$ surfaces present difficulties for local minimization
routines, the Markov chain Monte Carlo method is able to jump between
these local minima and accurately calculate the posterior probability
distribution for model parameters.  It is important to realize that if
there were multiple local minima separated by a barrier region of
parameter space which resulted in significantly larger values of
$\chi^2$, then Markov chain Monte Carlo would have great difficulty
sampling from the posterior distribution unless special care was taken
to identify the well separated local minima and a candidate transition
probability function was carefully chosen to allow transitions between
the local minima.  Thus, it is still important to conduct a single
global search over the parameter space to recognize if there are
distinct orbital solutions separated by a large $\chi^2(\vec{x})$
barrier.

In summary, MCMC has the following advantages compared to the other
methods we have discussed.
\begin{itemize}
\item MCMC naturally allows for correlated and non-Gaussian
uncertainties in fit parameters.
\item The results can be simply interpreted as the joint posterior
distribution for a given prior and set of observational data.
\item In some cases the resulting distribution can be efficiently
updated to account for additional observations or alternative priors,
provided that they do not greatly alter the posterior distribution.
\item Calculating the next step in a Markov chain is much faster than
performing an additional minimization with resampled data.
\item Computationally MCMC is more efficient than other techniques,
particularly for high dimensional parameter spaces.
\end{itemize}

\section{Example Applications}

To illustrate the application of the MCMC method to fitting radial
velocity data, we will present several example applications to the
presently known extrasolar planets.  For calculating the Markov chain
we use a set of parameters,
\begin{equation}
\vec{x} = ( \log P,\, \log K,\, e \cos \omega,\, e \sin \omega,\, M_o, \vec{C} ),
\label{EqnParam}
\end{equation}
where $\vec{C}$ is the set of mean velocity offsets (one offset for
each observatory).  It is common practice to choose a
``non-informative'' prior that is uniform in the logarithm of a
positive definite magnitude (e.g., orbital period and velocity
semi-amplitude), as suggested by several scaling arguments (Gelman et
al.\ 2003).  Additionally, we found that the use of $\log P$ and $\log
K$ instead of $P$ and $K$ increased the rate of convergence for
systems where the orbital period was not tightly constrained.
Similarly, we found that the use of $e \sin \omega$ and $e \cos
\omega$ instead of $e$ and $\omega$ significantly increased the rate
of convergence for systems with small eccentricities where $\omega$ is
not tightly constrained.

The distribution of published orbital parameters is roughly consistent
with being uniform in $\log P$, $\log K$, $e$, $\omega$, and $M_o$
(aside from cutoffs at small orbital periods and large planetary
masses and not including the very short period planets for which tidal
forces have presumablely circularized the orbits).  Therefore, for the
histograms and contour plots presented below, we have sampled from the
Markov chains using importance sampling so that they correspond to
priors which are flat on ($\log P$, $\log K$, $e$, $w$, $M_o$).

For each planetary system, we initialize multiple chains with
parameter values near the published values of the orbital elements or
the best-fit models identified with a GA.  We also start chains with
initial parameters drawn from a Gaussian distribution centered on the
published values with standard deviations three times the published
uncertainties.  In cases where our Markov chains indicated
uncertainties greater than the published values, we constructed
additional chains with initial parameters chosen over an even wider
range.  

For these simulations, we have used a Gaussian candidate transition
probability function (eqn.~ \ref{eqnCandTransProb}) and a Gibbs
sampler which randomly chooses to update one or two parameters at each
step with equal probability.  (When two parameters have significant
correlations, perturbing two parameters in a single step can allow the
Markov chain to explore parameter space more quickly.)  All parameters
other than the one or two chosen to be updated are left unchanged.
The order of the parameters to be altered is determined by a randomly
generated permutation.  We monitor the acceptance rates for steps
involving each parameter separately.  If a chain had an acceptance
rate below $0.25$ or above $0.55$, then we adjusted the relevant
$\beta_\mu$ and start computing a new chain.  To reduce the dependence
on the initial parameter values, we discard the initial $10000$ states
or the first 10\% of the chain, whichever is longer.  Since
consecutive states can are often highly correlated, we lose little
information by consider only every $m$-th set of parameter values with
$m$ greater than or equal to ten times the number of free parameters.

We test for non-convergence of each chain by comparing the
marginalized distribution of each parameter in the first half and full
chain.  We also compare the results of multiple chains run with
multiple values of $\vec{\beta}$.  This helps us recognize chains
where an inappropriate step size may have resulted in a chain becoming
trapped in a small region of parameter space for the duration of the
chain.
Additionally, we run multiple chains with widely dispersed initial
conditions to verify that $\sqrt{\hat{R}}$, the Gelman-Rubin test
statistic, is consistent with convergence.  Since we initialize the
chains with wide range of initial parameter values, it is easy to
recognize that the distributions of parameters based on only the early
portion of the chains are dependent on the values chosen for the
initial state of the Markov chain.  The Gelman-Rubin test statistics
can determine if the Markov chains have yet to converge by comparing
the variance of each parameter within each chain to the variance of
the parameter across multiple chains.  For Markov chains which
converge, the Gelman-Rubin test statistic approaches $1.0$ from above.
For our results we have checked that it is always less than $1.2$ and
usually less than $1.1$ (Gilks et al.\ 1996).  The plots presented are
based on only the Markov chain with the most desirable acceptance
rates.

These very conservative choices are computationally inefficient, but
the inefficiency can be tolerated for the purposes of this paper.  For
each system, we have computed tens of Markov chains, including a wide
range of $\left|\beta\right|$.  Each of our Markov chains typically
contains $10^6-10^{10}$ steps, depending on the number of fit
parameters and the choice of $\left|\beta\right|$.  (When
$\left|\beta\right|$ results in too high or low an acceptance rate,
convergence requires many more steps.)  Ideally, one would attempt to
improve the computational efficiency while maintaining confidence that
the chains have converged.  Based on our experience from the systems
studied in this paper, we believe that one could construct only five
Markov chains with a single value of $\left|\beta\right|$ (chosen to
give a desirable acceptance rate) for each system and still be
reasonablely confident that the Markov chains had converged.  We
believe that further refinements to the candidate transition
probability function could make MCMC even more efficient.  Using
additional changes of variables could reduce covariances between model
parameters could speed convergence and reduce the autocorrelation of
the states in each chain.  We hope to present such optimizations in a
future paper.

\subsection{Eccentricities}

Despite the nearly Gaussian uncertainties for the radial velocity
data, the uncertainties for orbital parameters can be significantly
non-Gaussian.  In particular, since the orbital eccentricity is
confined between zero and one, we might expect significant deviations
from normality when the eccentricity is small or nearly unity.  In
Fig.\ \ref{EWHistos} we show the distribution of eccentricities and
arguments of periastron for HD 76700, HD 68988, and HD 4203 based on
the published observations and uncertainties (Tinney et al.\ 2002;
Vogt et al.\ 2002).  There are several interesting features in these
plots:
\begin{enumerate}
\item For HD 76700, the distributions of orbital elements calculated
by resampling and MCMC are similar.  We find the
distribution of eccentricities barely includes $e=0$ (assuming a
uniform prior in $\vec{x}$ as defined in Eqn.\ (\ref{EqnParam})) and
excludes $e\le0.017$ at the 90\% confidence level, in 
contrast to the published solution, $e = 0\pm0.04$.  While the
published orbital solution does not constrain $\omega$, our
simulations constrain the argument of pericenter to
$-100^\circ\le\omega\le60^\circ$ (90\% confidence interval).
Additionally, the distribution of $\omega$ has a noticeable asymmetry.
For planets with small eccentricities such as HD 76700, the variables
$e \sin \omega$ and $e \cos \omega$ generally have smaller
uncertainties and are more Gaussian than the uncertainties in $e$ and
$\omega$.
\item For HD 68988, the eccentricity distribution is clearly separated
from zero, and the hypothesis that $e<0.1$ is strongly rejected within
the single planet Keplerian model.  Again, resampling and MCMC methods
result in similar uncertainty estimates.
\item HD 4203 has a significant eccentricity, however the published
observations still permit strongly non-Gaussian uncertainties in $e$.
In this case, MCMC results in an uncertainty distribution with a more
significant high eccentricity tail and a narrower distribution in
$\omega$ than is predicted by resampling.
\end{enumerate}

HD 80606 has a very large eccentricity ($e\simeq 0.93$) and a very
small radial separation at periastron ($r_p \simeq 0.035$AU).  While
the orbital period is very well constrained, $e$, and thus radial
separation at periastron ($r_p$) have significant uncertainties (Fig.\
\ref{RpHD80606}), especially in the context of tidal dissipation,
since the rate of tidal circularization is a steep function of $r_p$
\cite{RasToutLubLiv96}.  Our MCMC simulations show that the current
observations still permit orbital solutions with pericenter distances
as small as a stellar radius.  It is interesting to note that MCMC and
resampling methods result in significantly different distributions for
$r_p$.  The high correlation between variables does slow the
convergence of the Markov chains, but we have run several chains for
an extended duration and none result in a significant probability of
$r_p \ge 0.045$AU.  If additional observations were to constrain the
pericenter distance, this system might eventually provide an
interesting test for theories of tidal circularization and orbital
decay.

\subsection{Long Period Systems}

Constraining the orbital parameters of long period systems is a
significant challenge, since good solutions typically require
observations spanning at least one orbital period.  For eccentric
systems, it is much more difficult to identify optimal times to
observe the system while still during the first or second orbit.  In
Fig.\ \ref{PEContours} we show the joint distribution of the period
and eccentricity for the system HD 72659.  Note that the MCMC
simulations reveal a much larger uncertainty in orbital period and
eccentricity than suggested by resampling or by the published error
bars.  After this plot was prepared, Marcy (2003) informed us that
more recent observations of HD 72659 yield a best-fit of $P = 5828$ d
and $e=0.47$ (large solid point), demonstrating the potential for MCMC
to provide more accurate uncertainty estimates than resampling.

In Fig.\ \ref{ManyOrbitHistos} we show the marginalized probability
distributions for $P$, $K$, $e$, and $\omega$ for several of the known
extrasolar planets with the longest orbital periods.  While the
parameters are well constrained for some systems, significant
uncertainties in the periods and eccentricities appear to be common.
Both the mean value and uncertainty of orbital elements determined by
MCMC simulations can be significantly different from the values and
uncertainties suggested by resampling, particularly when at least one
parameter has a significantly skewed distribution.

\subsection{Multiple Orbital Solutions}

For some systems the observational data do not determine a unimodal
orbital solution; in particular there can be two similarly good fits
that are well separated from each other.  This is particularly common
in multiple planet systems.  In Fig.\ \ref{TwoSolutions47UMa} we show
two possible orbital solutions for 47 UMa.  One corresponds to the
published orbital solution and the other corresponds to a solution in
which the outer planet has a much longer orbital period and large
eccentricity.

We have begun conducting MCMC simulations to estimate the parameter
distributions for 47 UMa and other multiple planet systems.  For
multiple planet systems several orbital parameters can be highly
correlated greatly slowing convergence.  We are exploring changes of
variables which could accelerate convergence and permit more detailed
investigations of the uncertainties in the elements of such multiple
planet systems.

\subsection{Interacting Multiple Planet Systems}

The most easily observable deviation from Keplerian motion is generally
expected to be precession of the longitude of periastron.  This
suggests we adopt a precessing Keplerian model
\begin{equation}
\label{EqnPrecessingKeplerian}
\Delta v_{*,i}(t) = K_i \left[ \cos \left( T_i(t) + \omega_{o,i} + \dot{\omega}_i t \right) + e_i \cos \left( \omega_{o,i} + \dot{\omega}_i t \right) \right],
\end{equation}
for each planet, where $\dot{\omega}_i$ is the precession rate of the
longitude of periapse of the $i$th planet and $t$ is the time of the
observation.  Such a model incorporates the dominant observable
perturbation to the basic Keplerian model while introducing only one
additional parameter per planet.  This model is particularly good when
the perturbations excite a single dominant secular eigenmode, as in
GJ876, in which case the values of $\dot{\omega}_i$ should be nearly
equal and time-independent.  Further, in this model, $\dot{\omega}$
can be fit to radial velocity measurement without time consuming
N-body integrations.  Although radial velocity measurements constrain
only $m \sin i$, the precession rates can in principle determine the
planets' inclinations and masses.  Of course, this approach assumes
that there are no additional perturbations, for example due to general
relativistic effects, the quadrupole moment of the star, or additional
companions.

The GJ 876 system is an example of a interacting system in which
mutual perturbations are detectable.  Modeling GJ 876 with two precessing
Keplerian orbits and applying MCMC, we investigate the precession
rates of the two orbits.  In the top panel of Fig.\ \ref{OmegaDot}, we
show the probability distributions for $\dot{\omega}_1$ (dotted line),
$\dot{\omega}_2$ (dashed line), and the instantaneous average of the
two (solid line).  In the lower panel we plot the probability
distribution for the difference of the precession rates.  In this
case, the two precession rates are comparable, allowing for a secular
resonance in which $\Delta \omega$ remains small \cite{LeePeale2002}.  Based on the
published observations, we can not yet detect a significant difference
between the precession rates of the two planets.  Rates
differing by more than $\sim 10^\circ/\mathrm{yr}$ can be excluded
according to this model.

\subsection{Possible Transit Times}

Radial velocity measurements can be used to identify when a transit
could occur, if the system were to have a favorable inclination.  This
can greatly aid observers who can concentrate their efforts at the
appropriate times.  Correlations between best-fit orbital parameters
can have a significant impact on the search for transits.  Thus, it is
important that observers understand the uncertainty in the predicted
transit times.  Particularly, for strongly interacting multiple planet
systems, there can be significant uncertainties in the possible
transit times.

We have fit a model based on Eqn.\ (\ref{EqnPrecessingKeplerian}) to
the published radial velocity data for GJ876.  Using a genetic
algorithm we found a model with precession rates of $-44.0^o$/yr and
$-46.8^o$/yr for the inner and outer planets, respectively.  These are
similar to the precession rates found for the best-fit orbital
solution obtained via self consistent n-body integrations and local
minimization \cite{LaughCamb2001}.  Clearly, this comparison does not
include important short period perturbations to the orbits which
should be included when forecasting transits and planning
observations.  Additionally, a transit search would want to target the
time of ingress or egress which introduces additional dependences on
the stellar radii and orbital parameters.  Nevertheless, we use this
model to demonstrate the uncertainty in potential transit times in an
interacting system and the applicability of MCMC to predicting
potential transit times.  The actual system is most likely precessing
at a similar rate and the unknown precession rate contributes to the
uncertainty in predicting potential transit times.  In Fig.\
\ref{Transits} we show the probability distributions for the center of
the transit of GJ 876 b approximately two months after the last
published observation, assuming $\sin i = 1$.  The model which
includes the precession rates (solid line) allows for a significantly
wider range of transit times than the model which assumes fixed
periapses.  This significantly increases the duration over which GJ
876 must be monitored to ensure that the ingress or egress falls
within the window of observations.

\section{Conclusions}

Markov chain Monte Carlo methods provide a valuable tool for matching
planetary orbits to radial velocity data.  These techniques can be
particularly useful for characterizing the uncertainties and
correlations in estimated orbital parameters for extrasolar
planets.  Presently, several systems have large uncertainties in
orbital parameters, particularly the long period planets and multiple
planet systems.  Our Markov chain Monte Carlo simulations reveal
that the published orbital solutions can significantly
over- or underestimate the uncertainty in orbital parameters.
In addition to quantifying the uncertainties of various
orbital parameters, Markov chain Monte Carlo allows specific questions
to be investigated directly from the radial velocity data itself
bypassing fits to orbital parameters.  As more multiple planet systems
are discovered, such methods could become increasingly valuable.

\acknowledgments We thank Gilbert Holder and Hiranya Peiris for
discussions of Markov chain Monte Carlo techniques.  We thank Geoff
Marcy and Debra Fischer for discussions of observational issues and
sharing radial velocity observations before publication.  We thank
Gregory Laughlin for bringing to our attention the uncertainties in
possible transit times, and John Chambers for valuable discussions
about GJ 876.  This research was supported in part by NASA grant
NAG5-10456 and the EPIcS SIM Key Project.

\begin{figure}[ht]
\plotone{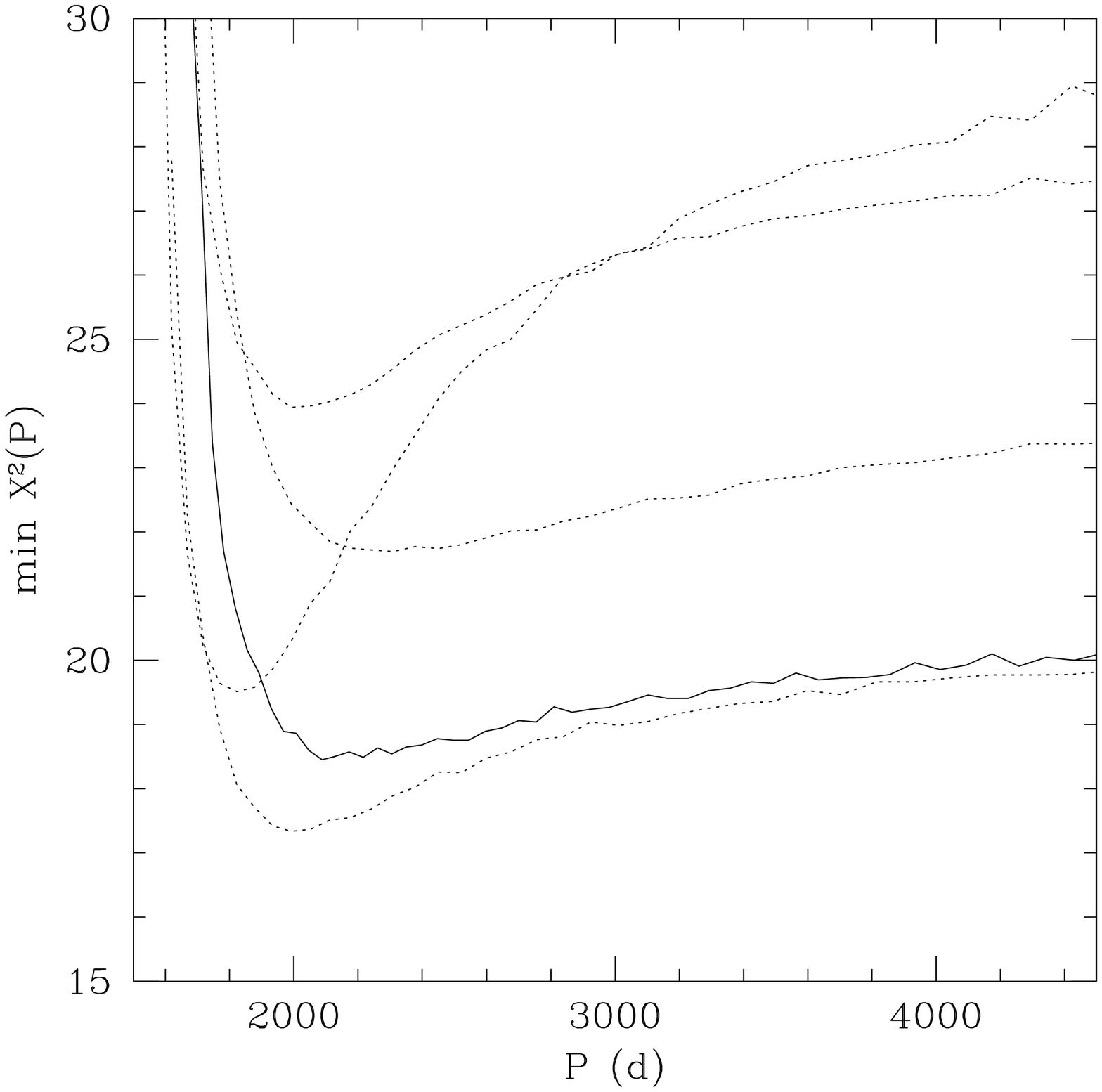}
\caption[Ford.fig1.ps]{ 
We show the minimum value of $\chi^2$ varrying all parameters except the orbital period
versus the orbital period for HD 72659.  The solid line is for the actual observations
and the dotted lines are based on resampled data.  Note that the very slow rise in $\chi^2(P)$ for
larger orbital periods allows a much greater range in the orbital period than suggested by 
the variation in the minimum of the $\chi^2(P)$ curves for resampled data.
\label{ChiSqVsPHD72659}}
\end{figure}

\begin{figure}[ht]
\plotone{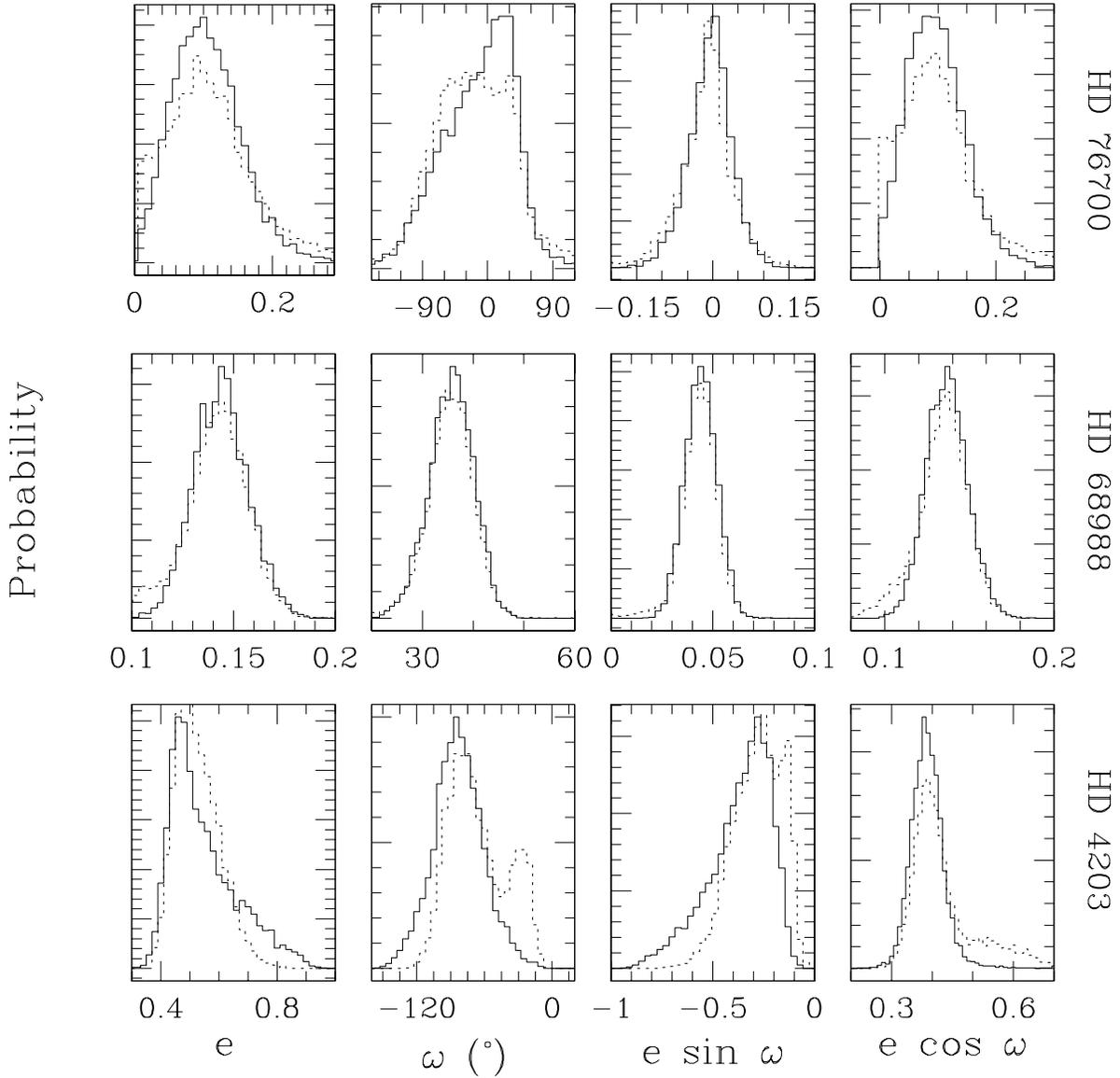}
\caption[Ford.fig2.ps]{ 
We show the distribution of eccentricities (far left), longitudes of
periastron (center left), and two related quantities, $e \sin \omega$
(center right) and $e \cos \omega$ (far right), for the companions to
HD 76700 (top), HD 68988 (middle), HD 4203 (bottom).
The solid lines represent the results
of our MCMC simulations, while the dotted lines show the results
of our fits to resampled data based on the published observations and
uncertainties (Tinney et al.\ 2002; Vogt et al.\ 2002).  
\label{EWHistos}}
\end{figure}

\begin{figure}[ht]
\plotone{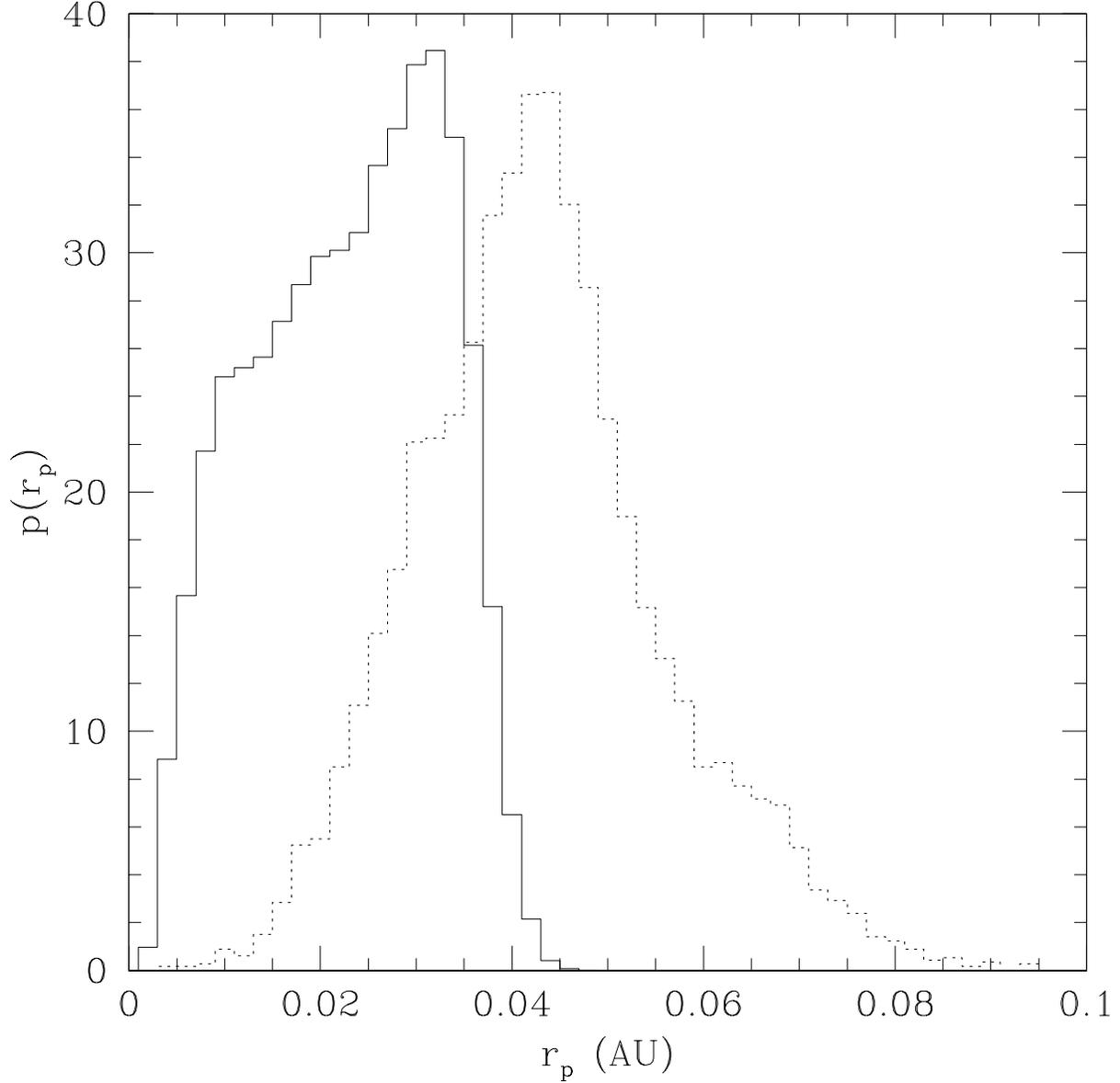}
\caption[Ford.fig3.ps]{ 
The probability distribution for the periastron distance of the
companion to HD 80606.  Here we assume a stellar mass of $1.1 M_\odot$
and inclinations corresponding to $\sin i =1$.  The solid line shows
the results of MCMC simulation and the dotted line shows the results
of fitting to resampled data.  Considering the strong dependence of
the tidal circularization rate on the periastron distance, the
uncertainty in the pericenter distance makes it impossible to
constrain theories of tidal circularization based on the present
observations.
\label{RpHD80606}}
\end{figure}

\begin{figure}[ht]
\plotone{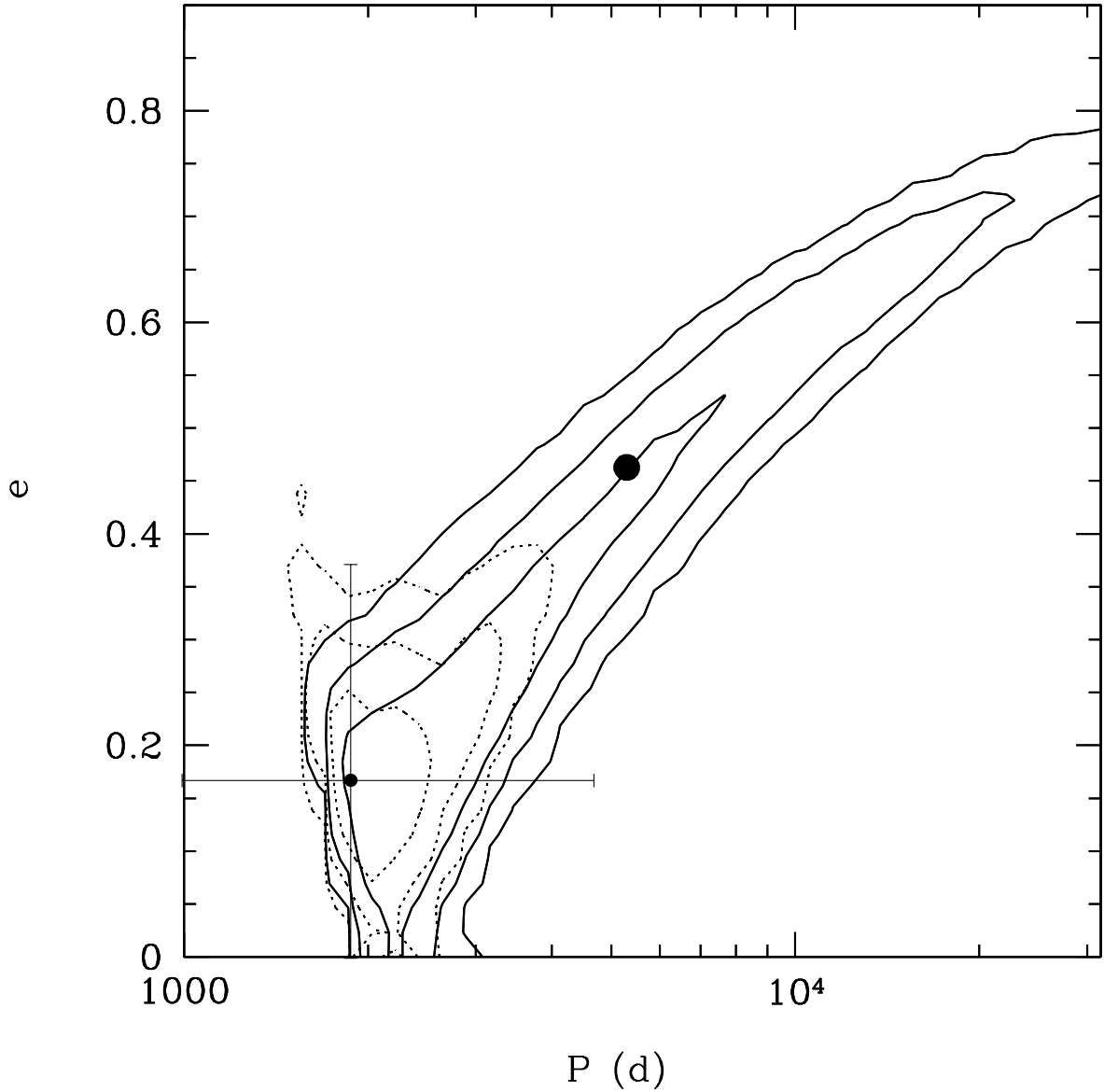}
\caption[Ford.fig4.ps]{ 
Here we show the probability distribution marginalized over all
variables except the period and eccentricity of HD 72659.  The solid
contours show the 1, 2, and 3-sigma confidence intervals, defined to
contain 68.3\%, 95.4\%, and 99.73\% of the probability distribution,
based on MCMC simulations.  The dotted contours also enclose the 1, 2,
and 3-sigma confidence intervals, but based on resampling methods.
The point with error bars indicates the published orbital solution and
uncertainties.  Both sets of contours assume a uniform prior in
$\log P$ and $e$.  The large point without error bars indicates the
best-fit orbital solution based on more recent observations made after
the rest of this plot had been prepared (Marcy 2003).  Unfortunately,
there are often large uncertainties in the orbital elements derived
for long period planets.  \\
\label{PEContours}}
\end{figure}

\begin{figure}[ht]
\plotone{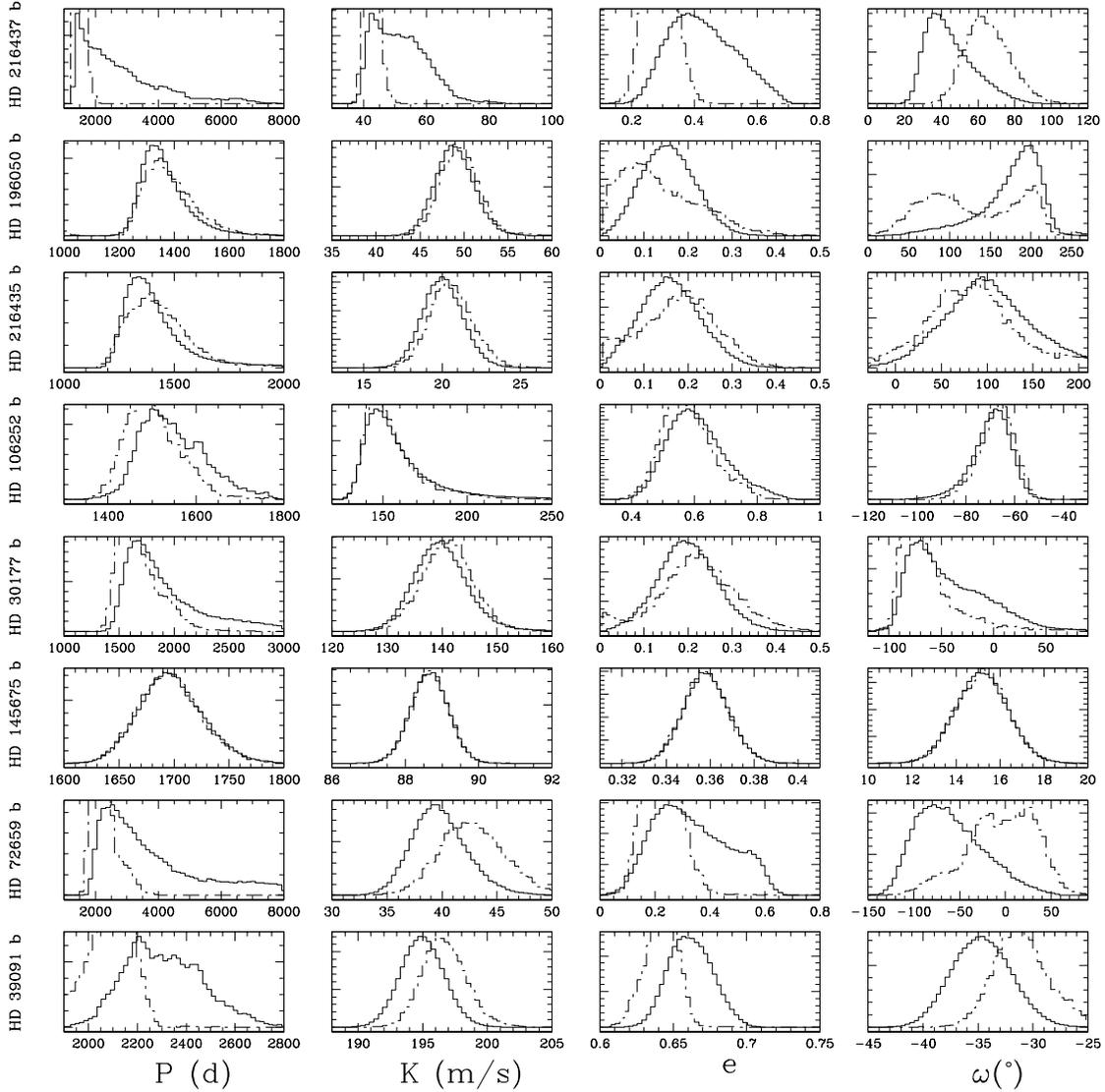}
\caption[Ford.fig5.ps]{ 
Here we present the probability distribution marginalized over all
but one variable for the eight extrasolar planets with the longest
orbital periods (we restrict ourselves to systems with only one known planet). 
The solid line shows the results of our MCMC simulations.  The dotted
line shows the results of our fits to resampled data.  The area under
each curve is normalized to unity, however the calculated
distributions sometimes have tails which extend beyond the domain of
these figures.
\label{ManyOrbitHistos}}
\end{figure}

\begin{figure}[ht]
\plotone{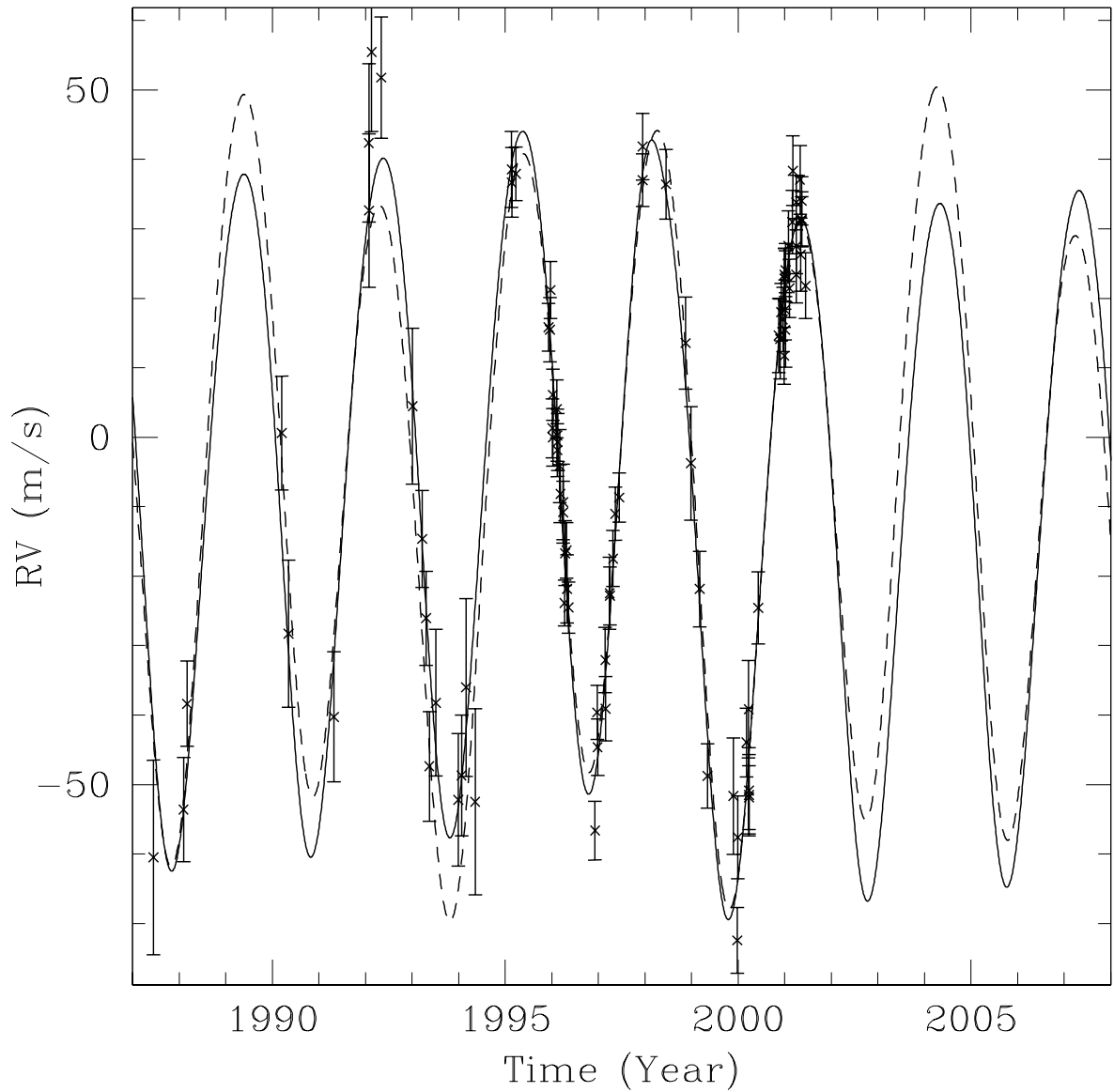}
\caption[Ford.fig6.ps]{ 
Here we present the published radial velocity data for 47 UMa along
with two fits to the data each employing two non-interacting planets
on Keplerian orbits.  The solid curve is the published solution, while
the dashed solution is for a much larger period (three times the
published best-fit period) and eccentricity ($e\simeq0.8$).  The
$\chi^2$ of the alternative fit is actually less than that of the
published fit, but the difference is not significant.  While the
dashed solution is not favored by orbital stability requirements, 47
UMa illustrates that the observations on multiple-planet systems can
sometimes be well modeled by very different orbital solutions.
\label{TwoSolutions47UMa}}
\end{figure}

\begin{figure}[ht]
\plotone{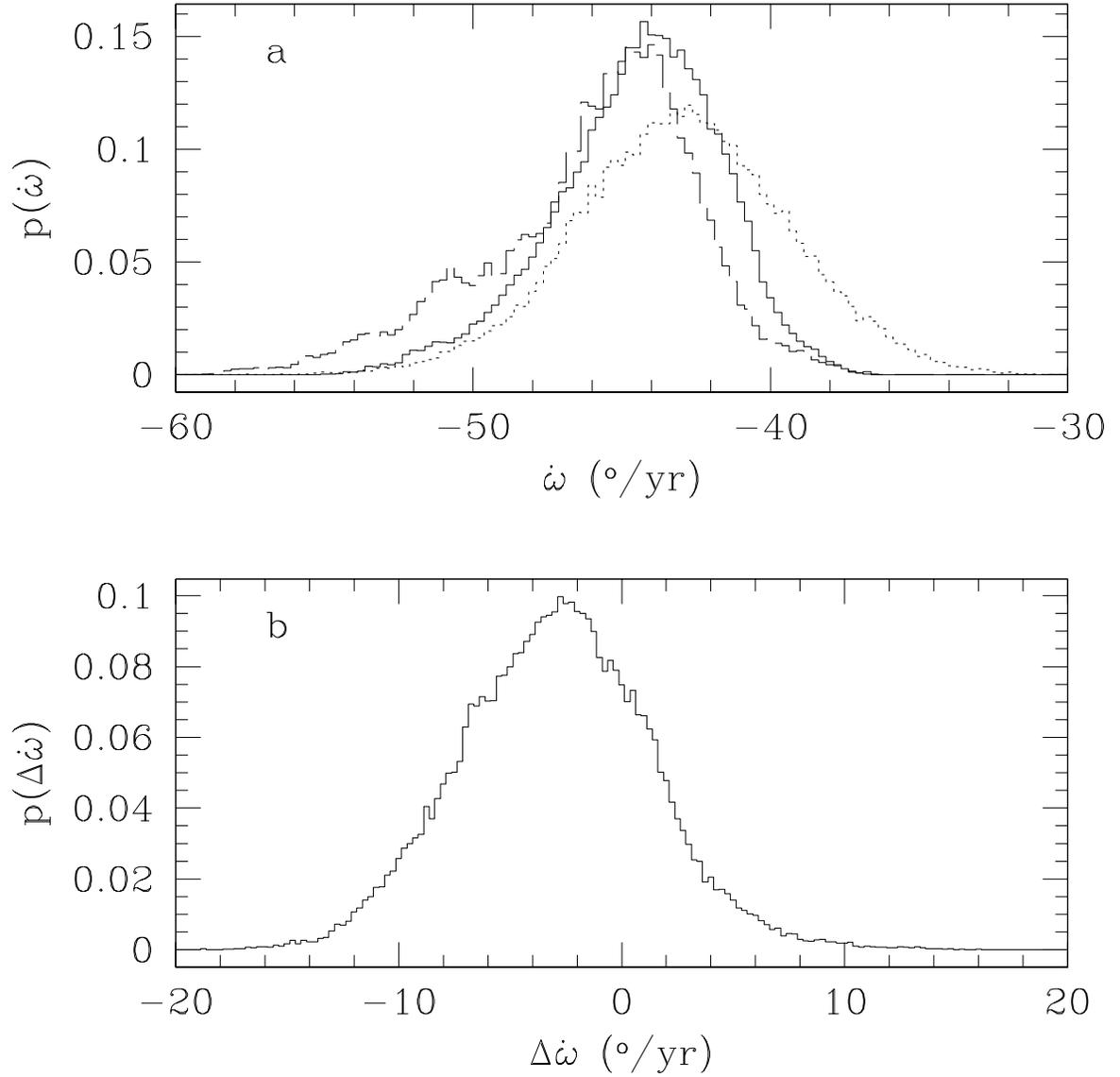}
\caption[Ford.fig7.ps]{ 
In the top panel we show the probability distributions for the
precession rates of the periapses of GJ 876 b (dotted) and c (dashed).
The solid line is the instantaneous average of the two precession
rates.  In the lower panel, we show the distribution for the
difference in the precession rates.
\label{OmegaDot}}
\end{figure}

\begin{figure}[ht]
\plotone{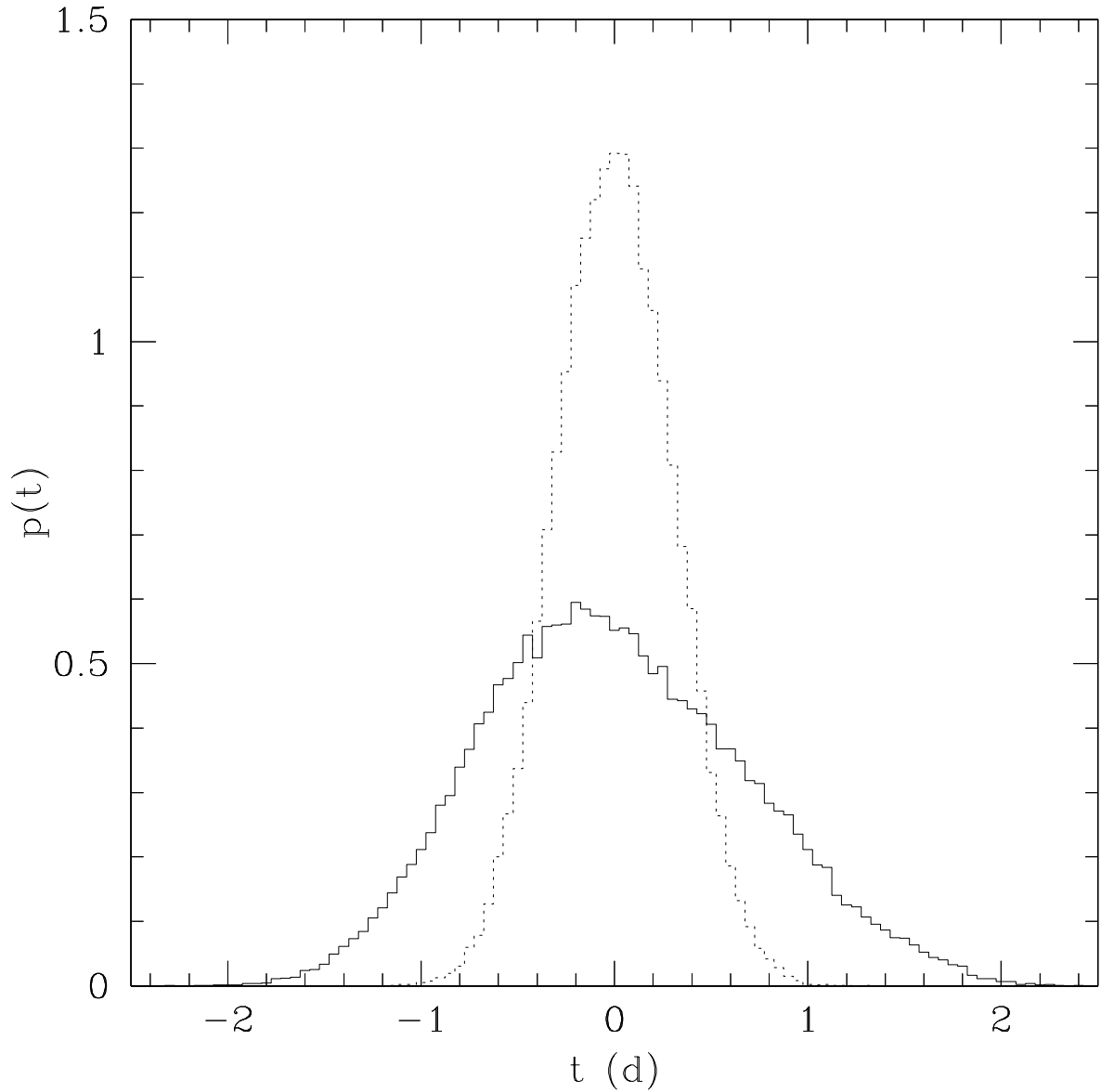}
\caption[Ford.fig8.ps]{ 
Here we show the times of the center of potential transits for GJ 876
b approximately two months after the last published radial velocity,
assuming $\sin i = 1$.  The dotted line shows the prediction using a
superposition of two non-interacting Keplerian orbits to model the
data, while the solid line allows each orbit to precess at an
arbitrary rate.  While our model does not include the short period
perturbations important for actually forecasting potential transits,
it does demonstrate the uncertainty of transit times in an interacting
system.
\label{Transits}}
\end{figure}

\end{document}